\documentclass[aps,twocolumn, floatfix, prx,longbibliography,superscriptaddress]{revtex4-1}
\usepackage{graphicx}
\DeclareGraphicsExtensions{.pdf}
\usepackage{amsmath,amssymb,bbold,bm,color}
\usepackage{float,hyperref}
\usepackage{epstopdf}
\usepackage{titlesec}
\hypersetup{colorlinks=true,urlcolor=blue}

\newcommand{\bk}{{\bm k}}

\newcommand{\bq}{{\bm q}}

\newcommand{\br}{{\bm r}}

\newcommand{\bj}{{\bm j}}
\newcommand{\bsig}{{\bm \sigma}}
\newcommand{\btau}{{\bm \tau}}

\newcommand{\cH}{{\cal H}}

\newcommand{\bee}{\begin{equation}}
\newcommand{\ee}{\end{equation}}

\begin{document}

\title{Proximitizing altermagnets with conventional superconductors}

\author{Niclas Heinsdorf}
\affiliation{Department of Physics and Astronomy, and Quantum Matter
  Institute, University of British Columbia, Vancouver, BC, Canada V6T 1Z1}
\affiliation{Max Planck Institute for Solid State Research, Heisenbergstrasse 1, 70569 Stuttgart, Germany} 

\author{Marcel Franz}
\affiliation{Department of Physics and Astronomy, and Quantum Matter
  Institute, University of British Columbia, Vancouver, BC, Canada V6T 1Z1}

\begin{abstract} 
Recent theoretical work highlighted unique properties of superconducting
altermagnets, including the wealth of topologically non-trivial phases as well as their potential
uses in spintronic applications. Given that no intrinsically
superconducting altermagnets have yet been discovered, we study here the
possibility of superconducting order induced  by proximity effect from
a conventional $s$-wave superconductor. Through symmetry analysis and
microscopic modeling we find that interesting superconducting phases
can indeed be proximity-induced in a thin altermagnetic film provided
that weak Rashba spin-orbit coupling is present at the
interface. Surprisingly, the resulting
superconductor is generically nodal with a 
mixed singlet/triplet order parameter and, importantly for
applications, capable of generating spin-polarized  persistent current. We propose a set of candidate heterostructures with low lattice mismatch suitable to probe these effects experimentally.

\end{abstract}

\date{\today}
\maketitle

{\em Introduction --} Spin-split electron bands, characteristic of
metallic altermagnets \cite{ahn2019antiferromagnetism,Hayami2020,Smejkal2022a,Smejkal2022b,Mazin2022}, make these novel quantum materials uniquely
suited for explorations into unconventional superconductivity. The key relevant
observation is that, with the exception of isolated points, electrons
on the Fermi surface of an altermagnet with momenta $\bk$ and $-\bk$
have the {\em same} spin $\sigma$ and are therefore incapable of
forming  conventional zero-momentum spin-singlet Cooper pairs. In the presence of
a weak attractive interaction the leading instability of a metallic
altermagnet, then, is either a uniform equal-spin {\em triplet} order
parameter \cite{Zhu2023,Heung2024, parshukov2025,Monkman2025}, or alternately a non-zero total momentum
Fulde-Ferrell-Larkin-Ovchinikov (FFLO) state \cite{Sumita2023,Chakraborty2024}.

A discovery of either of the above superconducting (SC)  phases would be rare and
interesting. The triplet order parameter, for instance, requires an
orbital pair wave function $g_{\bk}$ that is odd under inversion,
$g_{\bk}=-g_{-\bk}$. In simple models chiral $p$-wave order parameter with
$g_{\bk}\propto(k_x\pm i k_y)$ is often energetically favored
\cite{Zhu2023,Heung2024}. Such a 
spin-triplet chiral $p$-wave superconductor is known to be
topologically non-trivial with protected chiral edge states and
Majorana zero modes in vortex cores \cite{Read2000}. In addition, the specific variant
likely to occur in altermagnetic (ALM) metals has two independent order
parameters for two spin projections which makes it uniquely suited
for the generation and transport of persistent spin currents of interest
in spintronic applications \cite{Monkman2025}.     
   
Although several altermagnets that are also good metals have been
discovered over the past two years, none so far has been reported to
exhibit superconductivity. This motivates us to address the following
two questions: ($i$) Is it possible to induce superconductivity in a
metallic altermagnet by proximity effect with a conventional spin
singlet $s$-wave superconductor? ($ii$) If so, what is the nature of
the resulting SC state? These questions are nontrivial because it is
not apriori clear how would tunneling of spin-singlet Cooper pairs
generate SC order in an altermagnet whose susceptibility to
spin-singlet pairing vanishes. We note that Ref.\ \cite{Cano2024}
studied proximitized altermagnets and found numerous interesting
topological phases. However, this work simply assumed that SC order would
be generated and did not consider the underlying microscopic
mechanism, which is of primary importance if we wish to understand
conditions under which such phases might occur in realistic devices.  

\begin{figure}[t]
\includegraphics[width = 8.6cm]{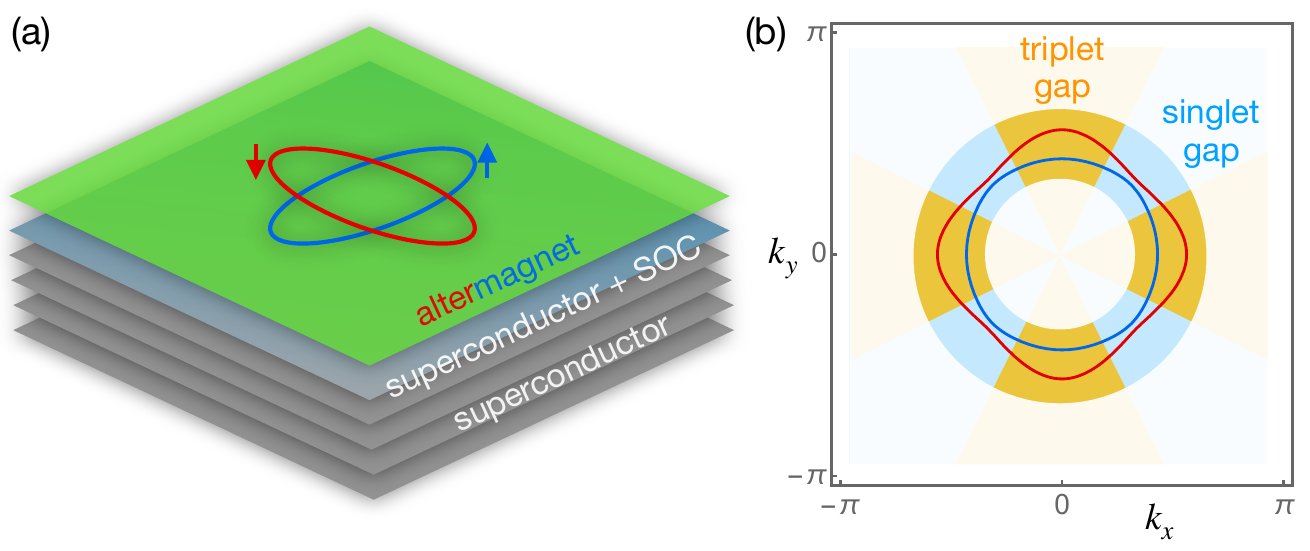}
\caption{(a) Sketch of the proposed setup: Altermagnetic normal metal (green)
  supported by a conventional $s$-wave superconductor (grey). The
  topmost layer of the SC substrate (blue-grey) is assumed to have Rashba SOC. (b)
  Schematic Fermi surface of the proximitized altermagnet. The induced
  Rashba SOC splits the spin degeneracy along the zone diagonal. The
  resulting FS is gapped out by a predominantly triplet or singlet SC
  order parameter as indicated, with Dirac nodal points associated with the boundaries.   
}. 
\label{fig1}
\end{figure}
To address the two questions posed above we consider a
simple setup shown in Fig.\ \ref{fig1}(a) consisting of a thin-film
ALM metal supported on 
a substrate made of a conventional superconductor such as Pb or Al.
We find that non-vanishing Rashba spin-orbit coupling (SOC) is required at the
interface if one wants to generate a SC state with a gap over most of
the altermagnet's Fermi surface. The resulting SC state is a mixture 
of spin singlet and triplet orders, which is symmetry-allowed in the
presence broken inversion symmetry. Specifically,  as indicated in
Fig.\ \ref{fig1}(b), we find a
proximity-induced state with portions of the FS near spin-degeneracy
points gapped by spin-singlet SC gap while the rest is dominated by
spin-triplet gap. The two types
of gapped regions are separated by nodal 
points with Dirac quasiparticle excitations, reminiscent of nodal
superconductivity found in high-$T_c$ cuprates \cite{Kirtley2000}. In the minimal model of
a $d$-wave altermagnet there are 8 nodal points in the
Brillouin zone (BZ), compared to 4 in cuprates. Nevertheless their presence
leads to similar low-temperature phenomenology, including the $T^2$
behavior of the electronic specific heat \cite{Moler1994} and $T$-linear superfluid
density \cite{Annett1991,Hardy1993}. Importantly, we find that parts
of the FS with induced spin-triplet SC gap 
are capable of supporting persistent {\em spin current generation} that
could be of interest for spintronic applications.

{\em Symmetry considerations --} At the level of phenomenological
Ginzburg-Landau (GL) theory the proximity effect arises from Cooper
pair tunneling from superconductor into normal
metal. Mathematically this is described by a term in the GL free energy density of the form
\begin{equation}\label{h1}
  f_{p}[\psi_0,\psi_1]=-D(\psi_0^\ast\psi_1+{\rm c.c.}),
  \end{equation}
where $\psi_{0/1}$ represent the complex scalar SC order parameters in
the superconductor and normal metal, respectively. In the
superconductor $|\psi_0|>0$, and the linear coupling in Eq.\
\eqref{h1} ensures that in equilibrium $\psi_1$ will be also
non-zero whenever $D\neq 0$. Of course $D$ will only be non-zero when
$f_{p}$ is invariant under all symmetry operations. For concreteness,
we focus below on a system with C$_4$ symmetry. 

When
$\psi_{0/1}$ both represent spin-singlet $s$-wave order parameters,
then the term in Eq.\ \eqref{h1} is allowed.
On the other hand, one might think that $D$ must
necessarily vanish if we take $\psi_1$ to represent a $p$-wave order parameter
appropriate for the ALM normal metal. (This is because
$\psi_1$ then transforms non-trivially under C$_4$ rotation,
$\psi_1\to i\psi_1$, whereas $\psi_0\to \psi_0$.)  However, this only
describes the orbital part of the pair wavefunction; in order to 
understand transformation properties of $\psi_1$ we must also consider
spin. As discussed in Refs.\ \cite{Heung2024,Monkman2025} the leading SC instability of a
$d$-wave altermagnet occurs in 4 degenerate channels whose full Cooper
pair wavefunctions can be written as
\begin{equation}\label{h2}
  p^\uparrow_\pm=|\uparrow\uparrow(p_x\pm ip_y)\rangle, \ \
   p^\downarrow_\pm=|\downarrow\downarrow(p_x\pm ip_y)\rangle.
\end{equation}
This expresses the simple fact that under equal-spin triplet pairing
there are two possible order parameter chiralities, labeled by $\pm$,
for each spin projection. How these order parameters transform under
2D rotations depends on the $z$-component of the total pair angular
momentum $J_z=S_z+L_z$, where $S_z$ and $L_z$ denote the spin and
orbital contribution, respectively. A quick thought reveals that
$p^\uparrow_-$ and  $p^\downarrow_+$ both have $J_z=0$ and therefore
transform  {\em trivially} under rotations. It follows, then, that a
linear coupling shown in Eq.\ \eqref{h1} is
allowed for these two order parameters when $\psi_0$ also transforms
trivially.

Importantly, the above is true when the inversion symmetry is also broken at the
surface ($p$-wave is odd under inversion whereas $s$-wave is even). In
the microscopic model considered next we will account for this by
including a Rashba SOC term at the SC/ALM interface which also
facilitates conversion of spin-singlet to spin-triplet Cooper
pairs. We conclude that, remarkably, symmetry considerations permit $p^\uparrow_-$
and  $p^\downarrow_+$ order parameters to be proximity-induced in the
altermagnet from an ordinary spin-singlet $s$-wave superconductor.

{\em Microscopic model --} To understand how this occurs at the level
of electron degrees of freedom, we turn to a minimal microscopic
model. Specifically, we model the proximity effect in the setup depicted in Fig.\
\ref{fig1}(a) by a lattice Hamiltonian $\cH=\cH_{\rm SC}+\cH_{\rm
  ALM}+\cH_{g}$ where
\begin{eqnarray}\label{h3}
\cH_{\rm SC}&=&
\sum_{\bk}\left[ \xi_\bk+2\lambda_R(\sigma^x\sin{k_y}-\sigma^y\sin{k_x})\right]_{\sigma\sigma'}
         d^\dagger_{\bk\sigma}d_{\bk\sigma'}\nonumber \\
&+&\Delta_0 \sum_{\bk} (d^\dagger_{\bk\uparrow}d^\dagger_{-\bk\downarrow} +{\rm h.c.})
\end{eqnarray}
describes the topmost layer of the SC substrate. Here,
$d^\dagger_{\bk\sigma}$ creates an electron with momentum $\bk$ and
spin $\sigma$ with $\bsig$ denoting the vector of
Pauli matrices in spin space and $\lambda_R$ the strength of Rashba
SOC. $\xi_\bk=-2t(\cos{k_x}+\cos{k_y})-\mu$ is the electron
dispersion referenced to the chemical potential $\mu$. We note that
the Rashba term is generically allowed since the inversion symmetry is
necessarily broken at the surface.

The altermagnet is described by the standard $d$-wave model
\cite{Smejkal2022a,Smejkal2022b,sitedecoupling1} on a square lattice
\begin{equation}\label{h4}
\cH_{\rm ALM}=
\sum_{\bk}[\xi'_\bk+\sigma^z\eta_\bk]_{\sigma\sigma'}c^\dagger_{\bk\sigma}c_{\bk\sigma'}
\end{equation}
with the altermagnetic
splitting captured by $\eta_\bk=2\eta_0(\cos{k_x}-\cos{k_y})$ and
$\xi'_\bk=-2t'(\cos{k_x}+\cos{k_y})-\mu'$. The two systems are
coupled by spin and momentum conserving tunneling,
\begin{equation}\label{h5}
\cH_{g}=
g\sum_{\bk}\left(c^\dagger_{\bk\sigma}d_{\bk\sigma}+{\rm h.c.}\right)
\end{equation}
with real amplitude $g$. We remark that it is not difficult to enlarge
the above model by, e.g.,\ including additional SC layers or through
lattice mismatch. However, such extensions do not bring any qualitatively new
features so we focus on the minimal model defined above. 

It will be useful to recast the Hamiltonian $\cH$ in the standard Nambu
matrix notation. The SC part can be written compactly as $\cH_{\rm SC}=
\sum_{\bk}\Phi_\bk^\dag h_0(\bk)\Phi_\bk$ with
$\Phi_\bk=(d_{\bk\uparrow},d_{\bk\downarrow},
d^\dagger_{-\bk\downarrow} ,-d^\dagger_{-\bk\uparrow})^T$ and
\begin{equation}\label{h6}
h_0(\bk)=\tau^z[\xi_\bk+2\lambda_R(\sigma^x\sin{k_y}-\sigma^y\sin{k_x})]+\tau^x\Delta_0.
\end{equation}
Here $\btau$ are Pauli matrices in the Nambu space.  Similarly, the
altermagnet can be represented as
\begin{equation}\label{h7}
h_1(\bk)=\tau^z\xi'_\bk+\sigma^z\eta_\bk.
\end{equation}
The full system is then described by an $8\times 8$ matrix Hamiltonian
\begin{equation}\label{h8}
  H_\bk=\begin{pmatrix}
    h_1(\bk) & V \\
    V^\dag  & h_0(\bk)
    \end{pmatrix},
\end{equation}
with $V=\tau^z g$.
\begin{figure*}[t]
\includegraphics[width = 17.6cm]{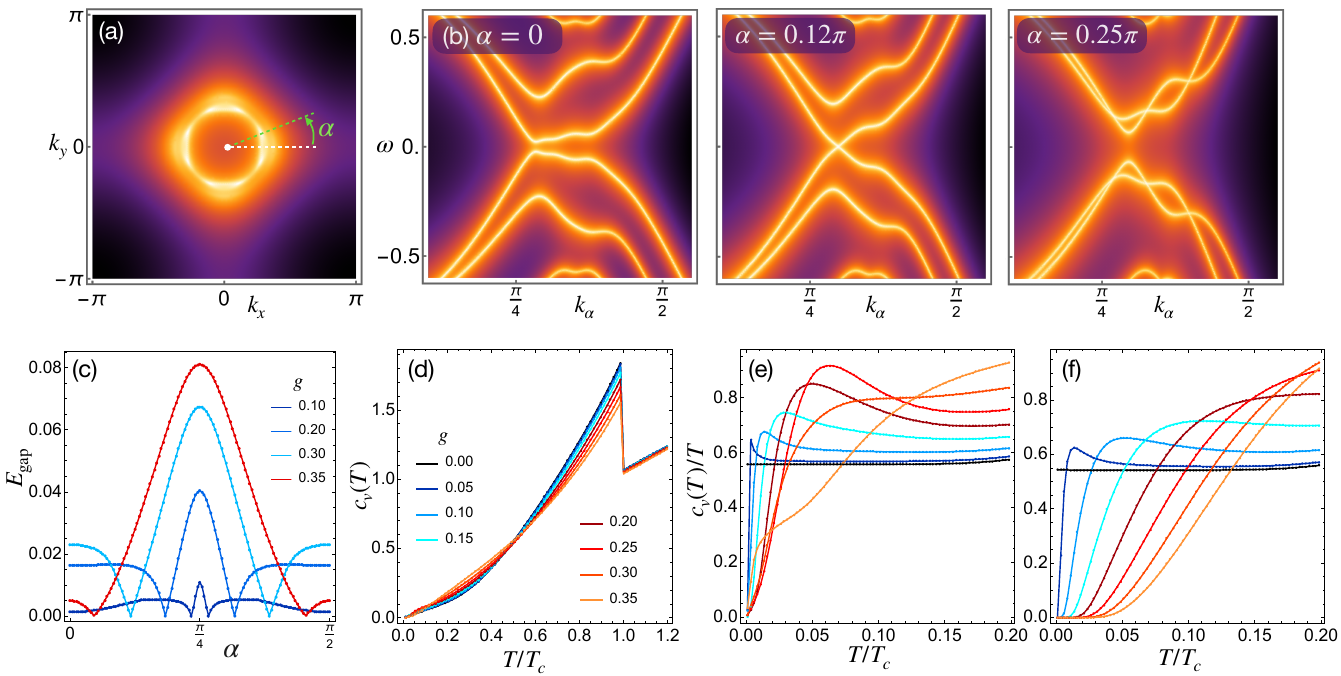}
\caption{Spectral properties of the proximitized altermagnet based on
  numerical diagonalization of Hamiltonian Eq.\ \eqref{h8} with
  parameters  $(t',\mu,\mu',\eta_0,\lambda_R,\Delta_0)=(1.0,-2.80,-3.05,
  0.2,0.1,0.4)$ and values of $g$ as indicated, all in units of $t$. (a)
  Spectral function $A(\bk,\omega)$ of the altermagnetic
  layer at $\omega=0$ and $g=0.3$, visualizing the underlying Fermi
  surface. (b) $A(\bk,\omega)$ along momentum space
  cuts $\bk=k_\alpha(\cos{\alpha},\sin{\alpha})$. Intensities are rendered on the logarithmic scale for better clarity.   (c)
  Excitation gap $E_{\rm gap}$ as a function of angle $\alpha$. (d)
  Specific heat $c_v(T)$ as a function of temperature $T$ and (e)
 detail of $c_v(T)/T$ in the low-$T$ regime. (f) The same as (e) except
 in the non-magnetic phase with $\eta_0=0$. In panels (d-f) we use an
 interpolation formula for the BCS gap $\Delta(T)=\Delta_0 {\rm Tanh}[1.74{\rm
   Re}\sqrt{T_c/T-1} ]$, accurate over the entire range.
}\label{fig2}
\end{figure*}

{\em Effective theory --} It is instructive to derive
an effective low-energy theory for the proximitized altermagnet by
formally integrating out the gapped electron degrees of freedom
residing in the SC layer. This is most easily achieved by performing a unitary transformation on $H_\bk$ that renders the
transformed Hamiltonian $\tilde{H}_\bk$ block diagonal. Following the
steps outlined in Ref.\ \cite{Fang2007}, we thus obtain, to second order in
coupling $g$,
\begin{equation}\label{h9}
\tilde{h}_{1}=h_{1}-{1\over 2}\left[Vh_{0}^{-1}V^\dag + h_{\rm 1}
  Vh_{0}^{-2}V^\dag+{\rm h.c.}\right] +O(V^4),
\end{equation}
where we suppressed the momentum argument to reduce clutter. The
leading correction to $h_1(\bk)$ has the form
\begin{equation}\label{h10}
  \delta h_{1}(\bk)=\begin{pmatrix}
    \Omega_\bk & \Sigma_\bk \\
    \Sigma_\bk^\dag & -\Omega_\bk
    \end{pmatrix}, 
  \end{equation}
  where $\Omega_\bk$ and $\Sigma_\bk$ are $2\times 2$
  matrices. $\Omega_\bk$ represents a correction to the normal-state
  dispersion of the altermagnet and contains a small off-diagonal
  Rashba term $\sim g^2\lambda_R/\Delta_0^2$ that acts to resolve the
  spin degeneracies as indicated in Fig.\ \ref{fig1}(b).  More importantly
\begin{equation}\label{h11}
  \Sigma_\bk={g^2\Delta_0\over \epsilon_+^2\epsilon_-^2}
  \begin{pmatrix}
    \xi_\bk^2 +|\lambda_\bk|^2+\Delta_0^2 & -2\xi_\bk\lambda_\bk \\
   -2\xi_\bk\lambda_\bk^\ast&  \xi_\bk^2 +|\lambda_\bk|^2+\Delta_0^2 
    \end{pmatrix}, 
  \end{equation}
describes the SC proximity effect. Here
$\lambda_\bk=2\lambda_R(\sin{k_y}+i\sin{k_x})$ and
$\epsilon_\pm=\sqrt{(\xi_\bk\pm |\lambda_\bk|)^2+\Delta_0^2}$ are
eigenvalues of $h_0(\bk)$.

Diagonal and off-diagonal elements of $\Sigma_\bk$ in Eq.\ \eqref{h11}
represent the spin-singlet  and spin-triplet  components of the
pairing matrix, respectively. The later have the correct odd-parity $p$-wave orbital 
character owing to the $\lambda_\bk\sim i\lambda_R(k_x-ik_y)$ factor
coming from the Rashba SOC. We also note that the proximity-induced triplet order
parameter has exactly the $p_-^\uparrow\otimes p_+^\downarrow$
structure anticipated on the basis of the symmetry analysis.

The spectrum of the effective Hamiltonian $\tilde{h}_1=h_1+\delta h_1$
is analyzed in End Matter. We find that the singlet
component of the proximity-induced order parameter, although generally
larger than the triplet component, is effective in opening a gap only
near the zone diagonals where the altermagnetic spin splitting is
small. Elsewhere, the gap is dominated by the triplet component and
the two types of the gap are separated by {\em nodal points} where the
excitations are gapless. This gives rise to the overall gap structure displayed in
Fig.\ \ref{fig1}(b).

{\em Numerical results --} To validate our conclusions based on the effective
theory we numerically solve the full $8\times 8$ Hamiltonian Eq.\
\eqref{h8}; results are summarized in Fig.\ \ref{fig2}. Panels
(a,b) show the spectral function $A(\bk,\omega) =-{\rm Tr}'   {\rm Im} (\omega
+i\delta-H_\bk)^{-1}$ of electrons in the altermagnet layer  where ${\rm
  Tr}'$ indicates trace over the 4 elements pertaining to the ALM layer.  Momentum
cuts of $A(\bk,\omega) $ reveal two distinct gapped regions on the Fermi surface
separated by a nodal point. In order to better visualize the
proximity-induced gap we show in  Fig.\ \ref{fig2}(c) the gap
magnitude $E_{\rm gap}$ as a function of the $k$-space angle
$\alpha$. This is obtained by numerically finding the smallest
positive eigenvalue of $H_\bk$ along the line
$\bk=k_\alpha(\cos{\alpha},\sin{\alpha})$ for fixed $\alpha$. We observe that the nodal
point lies close to the zone diagonal for small $g$, defining a narrow
range of predominantly singlet-type gap. The gap grows in magnitude
with $g$ as the nodal points move further apart. As discussed in End
Matter this behavior can be understood from the effective theory.         

From the knowledge of the energy eigenvalues, it is straightforward to
deduce any thermodynamic quantity of interest, including the specific heat
$c_v(T)$, which we display in  Fig.\ \ref{fig2}(d). At small $g$, and
not too low temperature, we
observe a metallic  $c_v(T)\sim T$ contribution from the
nearly-gapless ALM layer superimposed on the exponentially activated BCS
specific heat with a jump at the critical temperature $T_c$. The
low-$T$ behavior of  $c_v(T)/T$ shown in Fig.\ \ref{fig2}(e) indicates a
$c_v(T)\sim T^2$ dependence characteristic of a 2D superconductor with
point nodes. This contrasts with the non-magnetic limit $\eta_0=0$ shown in
Fig.\ \ref{fig2}(f); in this case, the
proximity effect opens up a full gap everywhere on the Fermi surface resulting in an exponentially activated $c_v(T)$.
\begin{figure}[t]
\includegraphics[width = 8.6cm]{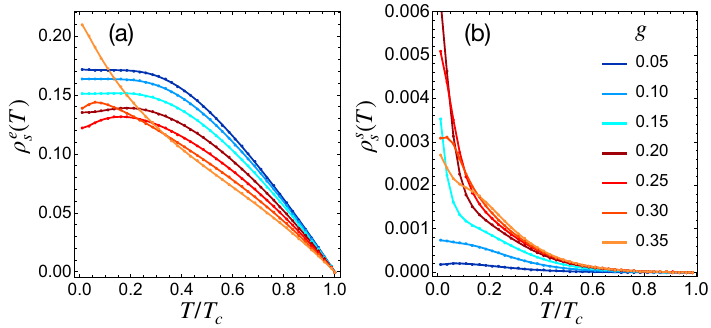}
\caption{Charge and spin superfluid densities of the combined bilayer
  system, defined in Eq.\ \eqref{h12}, plotted as a function of
  temperature $T$ for the same parameters as Fig.\ \ref{fig2}.    
} 
\label{fig3}
\end{figure}

{\em Persistent spin current generation --} Altermagnets with
intrinsic SC order have been predicted to support the spin-current dynamo
effect \cite{Monkman2025}, whereby charge supercurrent $\bj_e$ along the zone-diagonal direction
induces a pure spin supercurrent $\bj_s$ along the orthogonal direction with
magnitude $j_s \simeq (\eta_0/t) j_e$. To investigate this effect in
our bilayer device we set up charge supercurrent by imposing a phase gradient
$\Delta_0\to e^{-i\bq\cdot \br}\Delta_0$ on the Hamiltonian Eq.\ \eqref{h8} with
$\bq$ a vector along the $(1,1)$ direction. We then calculate
expectation values of both $\bj_e\parallel \bq$ and $\bj_s\perp\bq$
and quantify the bilayer 
response through its charge and spin superfluid densities, defined as
\begin{equation}\label{h12}
j_e=\rho_s^e q, \ \ \ j_s = \rho_s^sq. 
  \end{equation}

The behavior of the two superfluid densities as a function of 
temperature $T$ is shown in Fig.\ \ref{fig3}. For weak interlayer
coupling $g$ the charge superfluid 
density follows the standard BCS behavior with exponentially activated
suppression $\delta\rho_s^e(T)=\rho_s^e(T)-\rho_s^e(0)$  at low
$T$. Increasing $g$ initially causes a net decrease in 
$\rho_s^e(T)$ magnitude which we attribute to the pair-breaking effect of
the altermagnet on the superconductor (we checked that this
suppression does not occur in the non-magnetic limit
$\eta_0=0$). Eventually, at large enough 
$g$, this effect is balanced out by electrons in the altermagnet
contributing to the supercurrent leading to a net increase in
$\rho_s^e(0)$.  In this regime, we also observe $\delta\rho_s^e(T)\sim T$, a
behavior characteristic of nodal superconductors in the clean limit
\cite{Annett1991}, commonly observed in YBa$_2$Cu$_3$O$_{7-x}$
\cite{Hardy1993} and other clean high-$T_c$ cuprates \cite{Kirtley2000}.

Importantly, as shown in Fig.\ \ref{fig3}(b), the spin superfluid
density is non-zero at low $T$ confirming the 
presence of the spin-current dynamo effect. The magnitude of
$\rho_s^s$ grows as $g^2$ for small $g$ but the growth saturates when
$g\approx \Delta_0$ and $\rho_s^s$ never exceeds a few percent of
$\rho_s^e$. Nevertheless $j_s$ in this geometry represents pure spin supercurrent which could be technologically interesting.

{\em Conclusions --}  Contrary to the naive expectation, our results show
that it is possible to induce a spin-triplet SC order parameter in a
thin-film 
metallic altermagnet by the proximity effect from a conventional
spin-singlet $s$-wave superconductor, provided that Rashba-type
SOC is present at the interface. The resulting SC state in the altermagnet is
generically nodal with 8 Dirac nodes per BZ. The presence of Dirac nodes in the
quasiparticle excitation spectrum can be detected directly by
spectroscopic probes such as ARPES and STM, or through
thermodynamic quantities including specific heat $c_v(T)\sim T^2$
and superfluid density  $\delta\rho_s^e(T)\sim T$. In addition,
similar to the intrinsically superconducting altermagnets discussed
theoretically \cite{Monkman2025}, the proximitized altermagnet can be used to generate
persistent spin current. This includes pure spin supercurrent in the geometry where the phase
gradient points along the zone diagonal.

\begin{table}
\begin{center}
\begin{tabular}{ c c c c }
\hline
 SC & ALM & \ \ mismatch \ \ & \ \ symmetry ~ \ \ \\ \hline\hline
 Al & Rb$_{1-\delta}$V$_2$Te$_2$O & 0.04\% & tetragonal \\  
 NbS & FeSb$_2$ & 0.61\%, 3.21\% & orthorhombic \\ 
 Nb$_4$Se$_8$ & FeBr$_3$ & 2.23\% & hexagonal\\
  CaKFe$_4$As$_4$ & KV$_2$Se$_2$O & 2.87\% & tetragonal\\ 
 Pb & OsO$_2$ & 7.64\% & tetragonal\\ \hline
\end{tabular}
\end{center}
\caption{Proposed heterostructures of superconductors and ALM metals with favorable low lattice mismatches along the principal axes perpendicular to the stacking direction. Crystallographic data was taken from Refs.~\cite{jain2013commentary,haastrup2018computational,gjerding2021recent,ablimit2018weak,sodequist2024two,heinsdorf2021prediction,xu2025electronicstructuresmagnetictransition,raghuvanshi2025altermagnetic,mazin2021prediction,ruan2023superconductivity}. Note that spin–current dynamo effects are forbidden in heterostructures with hexagonal crystal symmetry. The mismatches for the orthorhombic case correspond to the $a$ and $b$ axis, respectively.}
\label{table:1}
\end{table}

Due to its nodal structure the proximity-induced SC state in our
model will exhibit flat band edge modes similar to those found in high-$T_c$ cuprates \cite{Lee2012}.
One could also imagine starting from an ALM normal state with spin-up and down Fermi pockets centered
around $X$ and $Y$ points of the BZ, respectively. When proximitized such a
system could become fully gapped and show more conventional topology with chiral or helical edge modes. 

Our conclusions follow from symmetry analysis and simple minimal
models and are therefore robust and broadly applicable. Table \ref{table:1} lists some candidate SC–ALM pairs with in-plane antiferromagnetic ordering wave vector and closely matched lattice constants~\cite{jain2013commentary,haastrup2018computational,gjerding2021recent,ablimit2018weak,sodequist2024two,heinsdorf2021prediction,xu2025electronicstructuresmagnetictransition,raghuvanshi2025altermagnetic,mazin2021prediction,ruan2023superconductivity}. In particular, a heterostructure formed of aluminum and the recently discovered oxychalcogenide altermagnet Rb$_{1-\delta}$V$_{2}$Te$_{2}$O shows a lattice mismatch of only $0.04\%$ and appears particularly promising~\cite{ablimit2018weak,FZhang2024}.  

Beyond experimental realization, several theoretical directions merit attention. These include the quantum geometry of the normal-state band structure, which has been shown to encode tendencies towards the formation of triplet superconductivity and altermagnetism~\cite{kitamura2024spin,sitedecoupling2}, and the role of the interface-induced Rashba SOC and related effects tied to the charge density profile across the heterostructure, which can be studied via {\em ab-initio} methods~\cite{yao2017manipulation}.


{\em Acknowledgments --} The authors are indebted to Yafis Barlas,
Ivar Martin, Benjamin T. Zhou and Stuart Parkin 
for stimulating discussions and correspondence. The work
was supported by NSERC, CIFAR and the Canada First Research Excellence Fund, Quantum Materials and Future Technologies Program. 
M.F.\ thanks Aspen Center for Physics where part of this work was completed.

\bibliography{spin}

\begin{thebibliography}{34}%
\makeatletter
\providecommand \@ifxundefined [1]{%
 \@ifx{#1\undefined}
}%
\providecommand \@ifnum [1]{%
 \ifnum #1\expandafter \@firstoftwo
 \else \expandafter \@secondoftwo
 \fi
}%
\providecommand \@ifx [1]{%
 \ifx #1\expandafter \@firstoftwo
 \else \expandafter \@secondoftwo
 \fi
}%
\providecommand \natexlab [1]{#1}%
\providecommand \enquote  [1]{``#1''}%
\providecommand \bibnamefont  [1]{#1}%
\providecommand \bibfnamefont [1]{#1}%
\providecommand \citenamefont [1]{#1}%
\providecommand \href@noop [0]{\@secondoftwo}%
\providecommand \href [0]{\begingroup \@sanitize@url \@href}%
\providecommand \@href[1]{\@@startlink{#1}\@@href}%
\providecommand \@@href[1]{\endgroup#1\@@endlink}%
\providecommand \@sanitize@url [0]{\catcode `\\12\catcode `\$12\catcode `\&12\catcode `\#12\catcode `\^12\catcode `\_12\catcode `\%12\relax}%
\providecommand \@@startlink[1]{}%
\providecommand \@@endlink[0]{}%
\providecommand \url  [0]{\begingroup\@sanitize@url \@url }%
\providecommand \@url [1]{\endgroup\@href {#1}{\urlprefix }}%
\providecommand \urlprefix  [0]{URL }%
\providecommand \Eprint [0]{\href }%
\providecommand \doibase [0]{http://dx.doi.org/}%
\providecommand \selectlanguage [0]{\@gobble}%
\providecommand \bibinfo  [0]{\@secondoftwo}%
\providecommand \bibfield  [0]{\@secondoftwo}%
\providecommand \translation [1]{[#1]}%
\providecommand \BibitemOpen [0]{}%
\providecommand \bibitemStop [0]{}%
\providecommand \bibitemNoStop [0]{.\EOS\space}%
\providecommand \EOS [0]{\spacefactor3000\relax}%
\providecommand \BibitemShut  [1]{\csname bibitem#1\endcsname}%
\let\auto@bib@innerbib\@empty
\bibitem [{\citenamefont {Ahn}\ \emph {et~al.}(2019)\citenamefont {Ahn}, \citenamefont {Hariki}, \citenamefont {Lee},\ and\ \citenamefont {Kune{\v{s}}}}]{ahn2019antiferromagnetism}%
  \BibitemOpen
  \bibfield  {author} {\bibinfo {author} {\bibfnamefont {Kyo-Hoon}\ \bibnamefont {Ahn}}, \bibinfo {author} {\bibfnamefont {Atsushi}\ \bibnamefont {Hariki}}, \bibinfo {author} {\bibfnamefont {Kwan-Woo}\ \bibnamefont {Lee}}, \ and\ \bibinfo {author} {\bibfnamefont {Jan}\ \bibnamefont {Kune{\v{s}}}},\ }\bibfield  {title} {\enquote {\bibinfo {title} {Antiferromagnetism in ruo 2 as d-wave pomeranchuk instability},}\ }\href {https://journals.aps.org/prb/abstract/10.1103/PhysRevB.99.184432} {\bibfield  {journal} {\bibinfo  {journal} {Physical Review B}\ }\textbf {\bibinfo {volume} {99}},\ \bibinfo {pages} {184432} (\bibinfo {year} {2019})}\BibitemShut {NoStop}%
\bibitem [{\citenamefont {Hayami}\ \emph {et~al.}(2019)\citenamefont {Hayami}, \citenamefont {Yanagi},\ and\ \citenamefont {Kusunose}}]{Hayami2020}%
  \BibitemOpen
  \bibfield  {author} {\bibinfo {author} {\bibfnamefont {Satoru}\ \bibnamefont {Hayami}}, \bibinfo {author} {\bibfnamefont {Yuki}\ \bibnamefont {Yanagi}}, \ and\ \bibinfo {author} {\bibfnamefont {Hiroaki}\ \bibnamefont {Kusunose}},\ }\bibfield  {title} {\enquote {\bibinfo {title} {Momentum-dependent spin splitting by collinear antiferromagnetic ordering},}\ }\href {https://journals.jps.jp/doi/10.7566/JPSJ.88.123702} {\bibfield  {journal} {\bibinfo  {journal} {journal of the physical society of japan}\ }\textbf {\bibinfo {volume} {88}},\ \bibinfo {pages} {123702} (\bibinfo {year} {2019})}\BibitemShut {NoStop}%
\bibitem [{\citenamefont {\ifmmode~\check{S}\else \v{S}\fi{}mejkal}\ \emph {et~al.}(2022{\natexlab{a}})\citenamefont {\ifmmode~\check{S}\else \v{S}\fi{}mejkal}, \citenamefont {Sinova},\ and\ \citenamefont {Jungwirth}}]{Smejkal2022a}%
  \BibitemOpen
  \bibfield  {author} {\bibinfo {author} {\bibfnamefont {Libor}\ \bibnamefont {\ifmmode~\check{S}\else \v{S}\fi{}mejkal}}, \bibinfo {author} {\bibfnamefont {Jairo}\ \bibnamefont {Sinova}}, \ and\ \bibinfo {author} {\bibfnamefont {Tomas}\ \bibnamefont {Jungwirth}},\ }\bibfield  {title} {\enquote {\bibinfo {title} {Beyond conventional ferromagnetism and antiferromagnetism: A phase with nonrelativistic spin and crystal rotation symmetry},}\ }\href {\doibase 10.1103/PhysRevX.12.031042} {\bibfield  {journal} {\bibinfo  {journal} {Phys. Rev. X}\ }\textbf {\bibinfo {volume} {12}},\ \bibinfo {pages} {031042} (\bibinfo {year} {2022}{\natexlab{a}})}\BibitemShut {NoStop}%
\bibitem [{\citenamefont {\ifmmode~\check{S}\else \v{S}\fi{}mejkal}\ \emph {et~al.}(2022{\natexlab{b}})\citenamefont {\ifmmode~\check{S}\else \v{S}\fi{}mejkal}, \citenamefont {Sinova},\ and\ \citenamefont {Jungwirth}}]{Smejkal2022b}%
  \BibitemOpen
  \bibfield  {author} {\bibinfo {author} {\bibfnamefont {Libor}\ \bibnamefont {\ifmmode~\check{S}\else \v{S}\fi{}mejkal}}, \bibinfo {author} {\bibfnamefont {Jairo}\ \bibnamefont {Sinova}}, \ and\ \bibinfo {author} {\bibfnamefont {Tomas}\ \bibnamefont {Jungwirth}},\ }\bibfield  {title} {\enquote {\bibinfo {title} {Emerging research landscape of altermagnetism},}\ }\href {\doibase 10.1103/PhysRevX.12.040501} {\bibfield  {journal} {\bibinfo  {journal} {Phys. Rev. X}\ }\textbf {\bibinfo {volume} {12}},\ \bibinfo {pages} {040501} (\bibinfo {year} {2022}{\natexlab{b}})}\BibitemShut {NoStop}%
\bibitem [{\citenamefont {Mazin}(2022)}]{Mazin2022}%
  \BibitemOpen
  \bibfield  {author} {\bibinfo {author} {\bibfnamefont {Igor}\ \bibnamefont {Mazin}} (\bibinfo {collaboration} {The PRX Editors}),\ }\bibfield  {title} {\enquote {\bibinfo {title} {Editorial: Altermagnetism---a new punch line of fundamental magnetism},}\ }\href {\doibase 10.1103/PhysRevX.12.040002} {\bibfield  {journal} {\bibinfo  {journal} {Phys. Rev. X}\ }\textbf {\bibinfo {volume} {12}},\ \bibinfo {pages} {040002} (\bibinfo {year} {2022})}\BibitemShut {NoStop}%
\bibitem [{\citenamefont {Zhu}\ \emph {et~al.}(2023)\citenamefont {Zhu}, \citenamefont {Zhuang}, \citenamefont {Wu},\ and\ \citenamefont {Yan}}]{Zhu2023}%
  \BibitemOpen
  \bibfield  {author} {\bibinfo {author} {\bibfnamefont {Di}~\bibnamefont {Zhu}}, \bibinfo {author} {\bibfnamefont {Zheng-Yang}\ \bibnamefont {Zhuang}}, \bibinfo {author} {\bibfnamefont {Zhigang}\ \bibnamefont {Wu}}, \ and\ \bibinfo {author} {\bibfnamefont {Zhongbo}\ \bibnamefont {Yan}},\ }\bibfield  {title} {\enquote {\bibinfo {title} {Topological superconductivity in two-dimensional altermagnetic metals},}\ }\href {\doibase 10.1103/PhysRevB.108.184505} {\bibfield  {journal} {\bibinfo  {journal} {Phys. Rev. B}\ }\textbf {\bibinfo {volume} {108}},\ \bibinfo {pages} {184505} (\bibinfo {year} {2023})}\BibitemShut {NoStop}%
\bibitem [{\citenamefont {Heung}\ and\ \citenamefont {Franz}(2025)}]{Heung2024}%
  \BibitemOpen
  \bibfield  {author} {\bibinfo {author} {\bibfnamefont {Tsz~Fung}\ \bibnamefont {Heung}}\ and\ \bibinfo {author} {\bibfnamefont {Marcel}\ \bibnamefont {Franz}},\ }\bibfield  {title} {\enquote {\bibinfo {title} {Probing topological degeneracy on a torus using superconducting altermagnets},}\ }\href {\doibase 10.1103/PhysRevB.111.205145} {\bibfield  {journal} {\bibinfo  {journal} {Phys. Rev. B}\ }\textbf {\bibinfo {volume} {111}},\ \bibinfo {pages} {205145} (\bibinfo {year} {2025})}\BibitemShut {NoStop}%
\bibitem [{\citenamefont {Parshukov}\ and\ \citenamefont {Schnyder}(2025)}]{parshukov2025}%
  \BibitemOpen
  \bibfield  {author} {\bibinfo {author} {\bibfnamefont {Kirill}\ \bibnamefont {Parshukov}}\ and\ \bibinfo {author} {\bibfnamefont {Andreas~P.}\ \bibnamefont {Schnyder}},\ }\href {https://arxiv.org/abs/2507.10700} {\enquote {\bibinfo {title} {Exotic superconducting states in altermagnets},}\ } (\bibinfo {year} {2025}),\ \Eprint {http://arxiv.org/abs/2507.10700} {arXiv:2507.10700 [cond-mat.supr-con]} \BibitemShut {NoStop}%
\bibitem [{\citenamefont {Monkman}\ \emph {et~al.}(2025)\citenamefont {Monkman}, \citenamefont {Weng}, \citenamefont {Heinsdorf}, \citenamefont {Nocera},\ and\ \citenamefont {Franz}}]{Monkman2025}%
  \BibitemOpen
  \bibfield  {author} {\bibinfo {author} {\bibfnamefont {Kyle}\ \bibnamefont {Monkman}}, \bibinfo {author} {\bibfnamefont {Joan}\ \bibnamefont {Weng}}, \bibinfo {author} {\bibfnamefont {Niclas}\ \bibnamefont {Heinsdorf}}, \bibinfo {author} {\bibfnamefont {Alberto}\ \bibnamefont {Nocera}}, \ and\ \bibinfo {author} {\bibfnamefont {Marcel}\ \bibnamefont {Franz}},\ }\href {https://arxiv.org/abs/2507.22139} {\enquote {\bibinfo {title} {Persistent spin currents in superconducting altermagnets},}\ } (\bibinfo {year} {2025}),\ \Eprint {http://arxiv.org/abs/2507.22139} {arXiv:2507.22139 [cond-mat.supr-con]} \BibitemShut {NoStop}%
\bibitem [{\citenamefont {Sumita}\ \emph {et~al.}(2023)\citenamefont {Sumita}, \citenamefont {Naka},\ and\ \citenamefont {Seo}}]{Sumita2023}%
  \BibitemOpen
  \bibfield  {author} {\bibinfo {author} {\bibfnamefont {Shuntaro}\ \bibnamefont {Sumita}}, \bibinfo {author} {\bibfnamefont {Makoto}\ \bibnamefont {Naka}}, \ and\ \bibinfo {author} {\bibfnamefont {Hitoshi}\ \bibnamefont {Seo}},\ }\bibfield  {title} {\enquote {\bibinfo {title} {Fulde-ferrell-larkin-ovchinnikov state induced by antiferromagnetic order in $\ensuremath{\kappa}$-type organic conductors},}\ }\href {\doibase 10.1103/PhysRevResearch.5.043171} {\bibfield  {journal} {\bibinfo  {journal} {Phys. Rev. Res.}\ }\textbf {\bibinfo {volume} {5}},\ \bibinfo {pages} {043171} (\bibinfo {year} {2023})}\BibitemShut {NoStop}%
\bibitem [{\citenamefont {Chakraborty}\ and\ \citenamefont {Black-Schaffer}(2024)}]{Chakraborty2024}%
  \BibitemOpen
  \bibfield  {author} {\bibinfo {author} {\bibfnamefont {Debmalya}\ \bibnamefont {Chakraborty}}\ and\ \bibinfo {author} {\bibfnamefont {Annica~M.}\ \bibnamefont {Black-Schaffer}},\ }\href {https://arxiv.org/abs/2309.14427} {\enquote {\bibinfo {title} {Zero-field finite-momentum and field-induced superconductivity in altermagnets},}\ } (\bibinfo {year} {2024}),\ \Eprint {http://arxiv.org/abs/2309.14427} {arXiv:2309.14427 [cond-mat.supr-con]} \BibitemShut {NoStop}%
\bibitem [{\citenamefont {Read}\ and\ \citenamefont {Green}(2000)}]{Read2000}%
  \BibitemOpen
  \bibfield  {author} {\bibinfo {author} {\bibfnamefont {N.}~\bibnamefont {Read}}\ and\ \bibinfo {author} {\bibfnamefont {Dmitry}\ \bibnamefont {Green}},\ }\bibfield  {title} {\enquote {\bibinfo {title} {Paired states of fermions in two dimensions with breaking of parity and time-reversal symmetries and the fractional quantum hall effect},}\ }\href {\doibase 10.1103/PhysRevB.61.10267} {\bibfield  {journal} {\bibinfo  {journal} {Phys. Rev. B}\ }\textbf {\bibinfo {volume} {61}},\ \bibinfo {pages} {10267--10297} (\bibinfo {year} {2000})}\BibitemShut {NoStop}%
\bibitem [{\citenamefont {Ghorashi}\ \emph {et~al.}(2024)\citenamefont {Ghorashi}, \citenamefont {Hughes},\ and\ \citenamefont {Cano}}]{Cano2024}%
  \BibitemOpen
  \bibfield  {author} {\bibinfo {author} {\bibfnamefont {Sayed Ali~Akbar}\ \bibnamefont {Ghorashi}}, \bibinfo {author} {\bibfnamefont {Taylor~L.}\ \bibnamefont {Hughes}}, \ and\ \bibinfo {author} {\bibfnamefont {Jennifer}\ \bibnamefont {Cano}},\ }\bibfield  {title} {\enquote {\bibinfo {title} {Altermagnetic routes to majorana modes in zero net magnetization},}\ }\href {\doibase 10.1103/PhysRevLett.133.106601} {\bibfield  {journal} {\bibinfo  {journal} {Phys. Rev. Lett.}\ }\textbf {\bibinfo {volume} {133}},\ \bibinfo {pages} {106601} (\bibinfo {year} {2024})}\BibitemShut {NoStop}%
\bibitem [{\citenamefont {Tsuei}\ and\ \citenamefont {Kirtley}(2000)}]{Kirtley2000}%
  \BibitemOpen
  \bibfield  {author} {\bibinfo {author} {\bibfnamefont {C.~C.}\ \bibnamefont {Tsuei}}\ and\ \bibinfo {author} {\bibfnamefont {J.~R.}\ \bibnamefont {Kirtley}},\ }\bibfield  {title} {\enquote {\bibinfo {title} {Pairing symmetry in cuprate superconductors},}\ }\href {\doibase 10.1103/RevModPhys.72.969} {\bibfield  {journal} {\bibinfo  {journal} {Rev. Mod. Phys.}\ }\textbf {\bibinfo {volume} {72}},\ \bibinfo {pages} {969--1016} (\bibinfo {year} {2000})}\BibitemShut {NoStop}%
\bibitem [{\citenamefont {Moler}\ \emph {et~al.}(1994)\citenamefont {Moler}, \citenamefont {Baar}, \citenamefont {Urbach}, \citenamefont {Liang}, \citenamefont {Hardy},\ and\ \citenamefont {Kapitulnik}}]{Moler1994}%
  \BibitemOpen
  \bibfield  {author} {\bibinfo {author} {\bibfnamefont {K.~A.}\ \bibnamefont {Moler}}, \bibinfo {author} {\bibfnamefont {D.~J.}\ \bibnamefont {Baar}}, \bibinfo {author} {\bibfnamefont {J.~S.}\ \bibnamefont {Urbach}}, \bibinfo {author} {\bibfnamefont {Ruixing}\ \bibnamefont {Liang}}, \bibinfo {author} {\bibfnamefont {W.~N.}\ \bibnamefont {Hardy}}, \ and\ \bibinfo {author} {\bibfnamefont {A.}~\bibnamefont {Kapitulnik}},\ }\bibfield  {title} {\enquote {\bibinfo {title} {Magnetic field dependence of the density of states of y${\mathrm{ba}}_{2}$${\mathrm{cu}}_{3}$${\mathrm{o}}_{6.95}$ as determined from the specific heat},}\ }\href {\doibase 10.1103/PhysRevLett.73.2744} {\bibfield  {journal} {\bibinfo  {journal} {Phys. Rev. Lett.}\ }\textbf {\bibinfo {volume} {73}},\ \bibinfo {pages} {2744--2747} (\bibinfo {year} {1994})}\BibitemShut {NoStop}%
\bibitem [{\citenamefont {Annett}\ \emph {et~al.}(1991)\citenamefont {Annett}, \citenamefont {Goldenfeld},\ and\ \citenamefont {Renn}}]{Annett1991}%
  \BibitemOpen
  \bibfield  {author} {\bibinfo {author} {\bibfnamefont {James}\ \bibnamefont {Annett}}, \bibinfo {author} {\bibfnamefont {Nigel}\ \bibnamefont {Goldenfeld}}, \ and\ \bibinfo {author} {\bibfnamefont {S.~R.}\ \bibnamefont {Renn}},\ }\bibfield  {title} {\enquote {\bibinfo {title} {Interpretation of the temperature dependence of the electromagnetic penetration depth in ${\mathrm{yba}}_{2}$${\mathrm{cu}}_{3}$${\mathrm{o}}_{7\mathrm{\ensuremath{-}}\mathrm{\ensuremath{\delta}}}$},}\ }\href {\doibase 10.1103/PhysRevB.43.2778} {\bibfield  {journal} {\bibinfo  {journal} {Phys. Rev. B}\ }\textbf {\bibinfo {volume} {43}},\ \bibinfo {pages} {2778--2782} (\bibinfo {year} {1991})}\BibitemShut {NoStop}%
\bibitem [{\citenamefont {Hardy}\ \emph {et~al.}(1993)\citenamefont {Hardy}, \citenamefont {Bonn}, \citenamefont {Morgan}, \citenamefont {Liang},\ and\ \citenamefont {Zhang}}]{Hardy1993}%
  \BibitemOpen
  \bibfield  {author} {\bibinfo {author} {\bibfnamefont {W.~N.}\ \bibnamefont {Hardy}}, \bibinfo {author} {\bibfnamefont {D.~A.}\ \bibnamefont {Bonn}}, \bibinfo {author} {\bibfnamefont {D.~C.}\ \bibnamefont {Morgan}}, \bibinfo {author} {\bibfnamefont {Ruixing}\ \bibnamefont {Liang}}, \ and\ \bibinfo {author} {\bibfnamefont {Kuan}\ \bibnamefont {Zhang}},\ }\bibfield  {title} {\enquote {\bibinfo {title} {Precision measurements of the temperature dependence of \ensuremath{\lambda} in ${\mathrm{yba}}_{2}$${\mathrm{cu}}_{3}$${\mathrm{o}}_{6.95}$: Strong evidence for nodes in the gap function},}\ }\href {\doibase 10.1103/PhysRevLett.70.3999} {\bibfield  {journal} {\bibinfo  {journal} {Phys. Rev. Lett.}\ }\textbf {\bibinfo {volume} {70}},\ \bibinfo {pages} {3999--4002} (\bibinfo {year} {1993})}\BibitemShut {NoStop}%
\bibitem [{\citenamefont {Roig}\ \emph {et~al.}(2024)\citenamefont {Roig}, \citenamefont {Kreisel}, \citenamefont {Yu}, \citenamefont {Andersen},\ and\ \citenamefont {Agterberg}}]{sitedecoupling1}%
  \BibitemOpen
  \bibfield  {author} {\bibinfo {author} {\bibfnamefont {Merc\`e}\ \bibnamefont {Roig}}, \bibinfo {author} {\bibfnamefont {Andreas}\ \bibnamefont {Kreisel}}, \bibinfo {author} {\bibfnamefont {Yue}\ \bibnamefont {Yu}}, \bibinfo {author} {\bibfnamefont {Brian~M.}\ \bibnamefont {Andersen}}, \ and\ \bibinfo {author} {\bibfnamefont {Daniel~F.}\ \bibnamefont {Agterberg}},\ }\bibfield  {title} {\enquote {\bibinfo {title} {Minimal models for altermagnetism},}\ }\href {\doibase 10.1103/PhysRevB.110.144412} {\bibfield  {journal} {\bibinfo  {journal} {Phys. Rev. B}\ }\textbf {\bibinfo {volume} {110}},\ \bibinfo {pages} {144412} (\bibinfo {year} {2024})}\BibitemShut {NoStop}%
\bibitem [{\citenamefont {Yao}\ \emph {et~al.}(2007)\citenamefont {Yao}, \citenamefont {Ye}, \citenamefont {Qi}, \citenamefont {Zhang},\ and\ \citenamefont {Fang}}]{Fang2007}%
  \BibitemOpen
  \bibfield  {author} {\bibinfo {author} {\bibfnamefont {Yugui}\ \bibnamefont {Yao}}, \bibinfo {author} {\bibfnamefont {Fei}\ \bibnamefont {Ye}}, \bibinfo {author} {\bibfnamefont {Xiao-Liang}\ \bibnamefont {Qi}}, \bibinfo {author} {\bibfnamefont {Shou-Cheng}\ \bibnamefont {Zhang}}, \ and\ \bibinfo {author} {\bibfnamefont {Zhong}\ \bibnamefont {Fang}},\ }\bibfield  {title} {\enquote {\bibinfo {title} {Spin-orbit gap of graphene: First-principles calculations},}\ }\href {\doibase 10.1103/PhysRevB.75.041401} {\bibfield  {journal} {\bibinfo  {journal} {Phys. Rev. B}\ }\textbf {\bibinfo {volume} {75}},\ \bibinfo {pages} {041401} (\bibinfo {year} {2007})}\BibitemShut {NoStop}%
\bibitem [{\citenamefont {Jain}\ \emph {et~al.}(2013)\citenamefont {Jain}, \citenamefont {Ong}, \citenamefont {Hautier}, \citenamefont {Chen}, \citenamefont {Richards}, \citenamefont {Dacek}, \citenamefont {Cholia}, \citenamefont {Gunter}, \citenamefont {Skinner}, \citenamefont {Ceder},\ and\ \citenamefont {Persson}}]{jain2013commentary}%
  \BibitemOpen
  \bibfield  {author} {\bibinfo {author} {\bibfnamefont {Anubhav}\ \bibnamefont {Jain}}, \bibinfo {author} {\bibfnamefont {Shyue~Ping}\ \bibnamefont {Ong}}, \bibinfo {author} {\bibfnamefont {Geoffroy}\ \bibnamefont {Hautier}}, \bibinfo {author} {\bibfnamefont {Wei}\ \bibnamefont {Chen}}, \bibinfo {author} {\bibfnamefont {William~Davidson}\ \bibnamefont {Richards}}, \bibinfo {author} {\bibfnamefont {Stephen}\ \bibnamefont {Dacek}}, \bibinfo {author} {\bibfnamefont {Shreyas}\ \bibnamefont {Cholia}}, \bibinfo {author} {\bibfnamefont {Dan}\ \bibnamefont {Gunter}}, \bibinfo {author} {\bibfnamefont {David}\ \bibnamefont {Skinner}}, \bibinfo {author} {\bibfnamefont {Gerbrand}\ \bibnamefont {Ceder}}, \ and\ \bibinfo {author} {\bibfnamefont {Kristin~A.}\ \bibnamefont {Persson}},\ }\bibfield  {title} {\enquote {\bibinfo {title} {Commentary: {{The Materials Project}}: {{A}} materials genome approach to accelerating materials innovation},}\ }\href {\doibase 10.1063/1.4812323} {\bibfield  {journal} {\bibinfo  {journal}
  {APL Materials}\ }\textbf {\bibinfo {volume} {1}},\ \bibinfo {pages} {011002} (\bibinfo {year} {2013})}\BibitemShut {NoStop}%
\bibitem [{\citenamefont {Haastrup}\ \emph {et~al.}(2018)\citenamefont {Haastrup}, \citenamefont {Strange}, \citenamefont {Pandey}, \citenamefont {Deilmann}, \citenamefont {Schmidt}, \citenamefont {Hinsche}, \citenamefont {Gjerding}, \citenamefont {Torelli}, \citenamefont {Larsen}, \citenamefont {Riis-Jensen} \emph {et~al.}}]{haastrup2018computational}%
  \BibitemOpen
  \bibfield  {author} {\bibinfo {author} {\bibfnamefont {Sten}\ \bibnamefont {Haastrup}}, \bibinfo {author} {\bibfnamefont {Mikkel}\ \bibnamefont {Strange}}, \bibinfo {author} {\bibfnamefont {Mohnish}\ \bibnamefont {Pandey}}, \bibinfo {author} {\bibfnamefont {Thorsten}\ \bibnamefont {Deilmann}}, \bibinfo {author} {\bibfnamefont {Per~S}\ \bibnamefont {Schmidt}}, \bibinfo {author} {\bibfnamefont {Nicki~F}\ \bibnamefont {Hinsche}}, \bibinfo {author} {\bibfnamefont {Morten~N}\ \bibnamefont {Gjerding}}, \bibinfo {author} {\bibfnamefont {Daniele}\ \bibnamefont {Torelli}}, \bibinfo {author} {\bibfnamefont {Peter~M}\ \bibnamefont {Larsen}}, \bibinfo {author} {\bibfnamefont {Anders~C}\ \bibnamefont {Riis-Jensen}},  \emph {et~al.},\ }\bibfield  {title} {\enquote {\bibinfo {title} {The computational 2d materials database: high-throughput modeling and discovery of atomically thin crystals},}\ }\href {https://iopscience.iop.org/article/10.1088/2053-1583/aacfc1} {\bibfield  {journal} {\bibinfo  {journal} {2D Materials}\
  }\textbf {\bibinfo {volume} {5}},\ \bibinfo {pages} {042002} (\bibinfo {year} {2018})}\BibitemShut {NoStop}%
\bibitem [{\citenamefont {Gjerding}\ \emph {et~al.}(2021)\citenamefont {Gjerding}, \citenamefont {Taghizadeh}, \citenamefont {Rasmussen}, \citenamefont {Ali}, \citenamefont {Bertoldo}, \citenamefont {Deilmann}, \citenamefont {Kn{\o}sgaard}, \citenamefont {Kruse}, \citenamefont {Larsen}, \citenamefont {Manti} \emph {et~al.}}]{gjerding2021recent}%
  \BibitemOpen
  \bibfield  {author} {\bibinfo {author} {\bibfnamefont {Morten~Niklas}\ \bibnamefont {Gjerding}}, \bibinfo {author} {\bibfnamefont {Alireza}\ \bibnamefont {Taghizadeh}}, \bibinfo {author} {\bibfnamefont {Asbj{\o}rn}\ \bibnamefont {Rasmussen}}, \bibinfo {author} {\bibfnamefont {Sajid}\ \bibnamefont {Ali}}, \bibinfo {author} {\bibfnamefont {Fabian}\ \bibnamefont {Bertoldo}}, \bibinfo {author} {\bibfnamefont {Thorsten}\ \bibnamefont {Deilmann}}, \bibinfo {author} {\bibfnamefont {Nikolaj~R{\o}rb{\ae}k}\ \bibnamefont {Kn{\o}sgaard}}, \bibinfo {author} {\bibfnamefont {Mads}\ \bibnamefont {Kruse}}, \bibinfo {author} {\bibfnamefont {Ask~Hjorth}\ \bibnamefont {Larsen}}, \bibinfo {author} {\bibfnamefont {Simone}\ \bibnamefont {Manti}},  \emph {et~al.},\ }\bibfield  {title} {\enquote {\bibinfo {title} {Recent progress of the computational 2d materials database (c2db)},}\ }\href {https://iopscience.iop.org/article/10.1088/2053-1583/ac1059/meta} {\bibfield  {journal} {\bibinfo  {journal} {2D Materials}\ }\textbf
  {\bibinfo {volume} {8}},\ \bibinfo {pages} {044002} (\bibinfo {year} {2021})}\BibitemShut {NoStop}%
\bibitem [{\citenamefont {Ablimit}\ \emph {et~al.}(2018)\citenamefont {Ablimit}, \citenamefont {Sun}, \citenamefont {Jiang}, \citenamefont {Wu}, \citenamefont {Liu},\ and\ \citenamefont {Cao}}]{ablimit2018weak}%
  \BibitemOpen
  \bibfield  {author} {\bibinfo {author} {\bibfnamefont {Abduweli}\ \bibnamefont {Ablimit}}, \bibinfo {author} {\bibfnamefont {Yun-Lei}\ \bibnamefont {Sun}}, \bibinfo {author} {\bibfnamefont {Hao}\ \bibnamefont {Jiang}}, \bibinfo {author} {\bibfnamefont {Si-Qi}\ \bibnamefont {Wu}}, \bibinfo {author} {\bibfnamefont {Ya-Bin}\ \bibnamefont {Liu}}, \ and\ \bibinfo {author} {\bibfnamefont {Guang-Han}\ \bibnamefont {Cao}},\ }\bibfield  {title} {\enquote {\bibinfo {title} {Weak metal-metal transition in the vanadium oxytelluride rb 1- $\delta$ v 2 te 2 o},}\ }\href {https://journals.aps.org/prb/abstract/10.1103/PhysRevB.97.214517} {\bibfield  {journal} {\bibinfo  {journal} {Physical Review B}\ }\textbf {\bibinfo {volume} {97}},\ \bibinfo {pages} {214517} (\bibinfo {year} {2018})}\BibitemShut {NoStop}%
\bibitem [{\citenamefont {S{\o}dequist}\ and\ \citenamefont {Olsen}(2024)}]{sodequist2024two}%
  \BibitemOpen
  \bibfield  {author} {\bibinfo {author} {\bibfnamefont {Joachim}\ \bibnamefont {S{\o}dequist}}\ and\ \bibinfo {author} {\bibfnamefont {Thomas}\ \bibnamefont {Olsen}},\ }\bibfield  {title} {\enquote {\bibinfo {title} {Two-dimensional altermagnets from high throughput computational screening: Symmetry requirements, chiral magnons, and spin-orbit effects},}\ }\href {https://pubs.aip.org/aip/apl/article-abstract/124/18/182409/3288014/Two-dimensional-altermagnets-from-high-throughput?redirectedFrom=fulltext} {\bibfield  {journal} {\bibinfo  {journal} {Applied Physics Letters}\ }\textbf {\bibinfo {volume} {124}} (\bibinfo {year} {2024})}\BibitemShut {NoStop}%
\bibitem [{\citenamefont {Heinsdorf}\ \emph {et~al.}(2021)\citenamefont {Heinsdorf}, \citenamefont {Christensen}, \citenamefont {Iraola}, \citenamefont {Zhang}, \citenamefont {Yang}, \citenamefont {Birol}, \citenamefont {Batista}, \citenamefont {Valent{\'\i}},\ and\ \citenamefont {Fernandes}}]{heinsdorf2021prediction}%
  \BibitemOpen
  \bibfield  {author} {\bibinfo {author} {\bibfnamefont {Niclas}\ \bibnamefont {Heinsdorf}}, \bibinfo {author} {\bibfnamefont {Morten~H}\ \bibnamefont {Christensen}}, \bibinfo {author} {\bibfnamefont {Mikel}\ \bibnamefont {Iraola}}, \bibinfo {author} {\bibfnamefont {Shang-Shun}\ \bibnamefont {Zhang}}, \bibinfo {author} {\bibfnamefont {Fan}\ \bibnamefont {Yang}}, \bibinfo {author} {\bibfnamefont {Turan}\ \bibnamefont {Birol}}, \bibinfo {author} {\bibfnamefont {Cristian~D}\ \bibnamefont {Batista}}, \bibinfo {author} {\bibfnamefont {Roser}\ \bibnamefont {Valent{\'\i}}}, \ and\ \bibinfo {author} {\bibfnamefont {Rafael~M}\ \bibnamefont {Fernandes}},\ }\bibfield  {title} {\enquote {\bibinfo {title} {Prediction of double-weyl points in the iron-based superconductor ca k fe 4 as 4},}\ }\href {https://journals.aps.org/prb/abstract/10.1103/PhysRevB.104.075101} {\bibfield  {journal} {\bibinfo  {journal} {Physical Review B}\ }\textbf {\bibinfo {volume} {104}},\ \bibinfo {pages} {075101} (\bibinfo {year}
  {2021})}\BibitemShut {NoStop}%
\bibitem [{\citenamefont {Xu}\ \emph {et~al.}(2025)\citenamefont {Xu}, \citenamefont {Zhang}, \citenamefont {Feng},\ and\ \citenamefont {Tian}}]{xu2025electronicstructuresmagnetictransition}%
  \BibitemOpen
  \bibfield  {author} {\bibinfo {author} {\bibfnamefont {Yuanji}\ \bibnamefont {Xu}}, \bibinfo {author} {\bibfnamefont {Huiyuan}\ \bibnamefont {Zhang}}, \bibinfo {author} {\bibfnamefont {Maoyuan}\ \bibnamefont {Feng}}, \ and\ \bibinfo {author} {\bibfnamefont {Fuyang}\ \bibnamefont {Tian}},\ }\href {https://arxiv.org/abs/2506.20968} {\enquote {\bibinfo {title} {The electronic structures, magnetic transition and fermi surface instability of room-temperature altermagnet kv$_{2}$se$_{2}$o},}\ } (\bibinfo {year} {2025}),\ \Eprint {http://arxiv.org/abs/2506.20968} {arXiv:2506.20968 [cond-mat.str-el]} \BibitemShut {NoStop}%
\bibitem [{\citenamefont {Raghuvanshi}\ \emph {et~al.}(2025)\citenamefont {Raghuvanshi}, \citenamefont {Berlijn}, \citenamefont {Parker}, \citenamefont {Wang}, \citenamefont {Manley}, \citenamefont {Hermann}, \citenamefont {Lindsay},\ and\ \citenamefont {Cooper}}]{raghuvanshi2025altermagnetic}%
  \BibitemOpen
  \bibfield  {author} {\bibinfo {author} {\bibfnamefont {Parul~R}\ \bibnamefont {Raghuvanshi}}, \bibinfo {author} {\bibfnamefont {Tom}\ \bibnamefont {Berlijn}}, \bibinfo {author} {\bibfnamefont {David~S}\ \bibnamefont {Parker}}, \bibinfo {author} {\bibfnamefont {Shaofei}\ \bibnamefont {Wang}}, \bibinfo {author} {\bibfnamefont {Michael~E}\ \bibnamefont {Manley}}, \bibinfo {author} {\bibfnamefont {Rapha{\"e}l~P}\ \bibnamefont {Hermann}}, \bibinfo {author} {\bibfnamefont {Lucas}\ \bibnamefont {Lindsay}}, \ and\ \bibinfo {author} {\bibfnamefont {Valentino~R}\ \bibnamefont {Cooper}},\ }\bibfield  {title} {\enquote {\bibinfo {title} {Altermagnetic behavior in oso 2: Parallels with ruo 2},}\ }\href {https://journals.aps.org/prmaterials/abstract/10.1103/PhysRevMaterials.9.034407} {\bibfield  {journal} {\bibinfo  {journal} {Physical Review Materials}\ }\textbf {\bibinfo {volume} {9}},\ \bibinfo {pages} {034407} (\bibinfo {year} {2025})}\BibitemShut {NoStop}%
\bibitem [{\citenamefont {Mazin}\ \emph {et~al.}(2021)\citenamefont {Mazin}, \citenamefont {Koepernik}, \citenamefont {Johannes}, \citenamefont {Gonz{\'a}lez-Hern{\'a}ndez},\ and\ \citenamefont {{\v{S}}mejkal}}]{mazin2021prediction}%
  \BibitemOpen
  \bibfield  {author} {\bibinfo {author} {\bibfnamefont {Igor~I}\ \bibnamefont {Mazin}}, \bibinfo {author} {\bibfnamefont {Klaus}\ \bibnamefont {Koepernik}}, \bibinfo {author} {\bibfnamefont {Michelle~D}\ \bibnamefont {Johannes}}, \bibinfo {author} {\bibfnamefont {Rafael}\ \bibnamefont {Gonz{\'a}lez-Hern{\'a}ndez}}, \ and\ \bibinfo {author} {\bibfnamefont {Libor}\ \bibnamefont {{\v{S}}mejkal}},\ }\bibfield  {title} {\enquote {\bibinfo {title} {Prediction of unconventional magnetism in doped fesb2},}\ }\href {https://www.pnas.org/doi/10.1073/pnas.2108924118} {\bibfield  {journal} {\bibinfo  {journal} {Proceedings of the National Academy of Sciences}\ }\textbf {\bibinfo {volume} {118}},\ \bibinfo {pages} {e2108924118} (\bibinfo {year} {2021})}\BibitemShut {NoStop}%
\bibitem [{\citenamefont {Ruan}\ \emph {et~al.}(2023)\citenamefont {Ruan}, \citenamefont {Yi}, \citenamefont {Chen}, \citenamefont {Zhou}, \citenamefont {Shi}, \citenamefont {Yang}, \citenamefont {Gu}, \citenamefont {Chen},\ and\ \citenamefont {Ren}}]{ruan2023superconductivity}%
  \BibitemOpen
  \bibfield  {author} {\bibinfo {author} {\bibfnamefont {Bin-Bin}\ \bibnamefont {Ruan}}, \bibinfo {author} {\bibfnamefont {Jun-Kun}\ \bibnamefont {Yi}}, \bibinfo {author} {\bibfnamefont {Le-Wei}\ \bibnamefont {Chen}}, \bibinfo {author} {\bibfnamefont {Menghu}\ \bibnamefont {Zhou}}, \bibinfo {author} {\bibfnamefont {Yun-Qing}\ \bibnamefont {Shi}}, \bibinfo {author} {\bibfnamefont {Qing-Song}\ \bibnamefont {Yang}}, \bibinfo {author} {\bibfnamefont {Ya-Dong}\ \bibnamefont {Gu}}, \bibinfo {author} {\bibfnamefont {Gen-Fu}\ \bibnamefont {Chen}}, \ and\ \bibinfo {author} {\bibfnamefont {Zhi-An}\ \bibnamefont {Ren}},\ }\bibfield  {title} {\enquote {\bibinfo {title} {Superconductivity in orthorhombic nbs},}\ }\href {https://journals.aps.org/prb/abstract/10.1103/PhysRevB.108.174517} {\bibfield  {journal} {\bibinfo  {journal} {Physical Review B}\ }\textbf {\bibinfo {volume} {108}},\ \bibinfo {pages} {174517} (\bibinfo {year} {2023})}\BibitemShut {NoStop}%
\bibitem [{\citenamefont {Wang}\ and\ \citenamefont {Lee}(2012)}]{Lee2012}%
  \BibitemOpen
  \bibfield  {author} {\bibinfo {author} {\bibfnamefont {Fa}~\bibnamefont {Wang}}\ and\ \bibinfo {author} {\bibfnamefont {Dung-Hai}\ \bibnamefont {Lee}},\ }\bibfield  {title} {\enquote {\bibinfo {title} {Topological relation between bulk gap nodes and surface bound states: Application to iron-based superconductors},}\ }\href {\doibase 10.1103/PhysRevB.86.094512} {\bibfield  {journal} {\bibinfo  {journal} {Phys. Rev. B}\ }\textbf {\bibinfo {volume} {86}},\ \bibinfo {pages} {094512} (\bibinfo {year} {2012})}\BibitemShut {NoStop}%
\bibitem [{\citenamefont {Zhang}\ \emph {et~al.}(2024)\citenamefont {Zhang}, \citenamefont {Cheng}, \citenamefont {Yin}, \citenamefont {Liu}, \citenamefont {Deng}, \citenamefont {Qiao}, \citenamefont {Shi}, \citenamefont {Zhang}, \citenamefont {Lin}, \citenamefont {Liu}, \citenamefont {Ye}, \citenamefont {Huang}, \citenamefont {Meng}, \citenamefont {Zhang}, \citenamefont {Okuda}, \citenamefont {Shimada}, \citenamefont {Cui}, \citenamefont {Zhao}, \citenamefont {Cao}, \citenamefont {Qiao}, \citenamefont {Liu},\ and\ \citenamefont {Chen}}]{FZhang2024}%
  \BibitemOpen
  \bibfield  {author} {\bibinfo {author} {\bibfnamefont {Fayuan}\ \bibnamefont {Zhang}}, \bibinfo {author} {\bibfnamefont {Xingkai}\ \bibnamefont {Cheng}}, \bibinfo {author} {\bibfnamefont {Zhouyi}\ \bibnamefont {Yin}}, \bibinfo {author} {\bibfnamefont {Changchao}\ \bibnamefont {Liu}}, \bibinfo {author} {\bibfnamefont {Liwei}\ \bibnamefont {Deng}}, \bibinfo {author} {\bibfnamefont {Yuxi}\ \bibnamefont {Qiao}}, \bibinfo {author} {\bibfnamefont {Zheng}\ \bibnamefont {Shi}}, \bibinfo {author} {\bibfnamefont {Shuxuan}\ \bibnamefont {Zhang}}, \bibinfo {author} {\bibfnamefont {Junhao}\ \bibnamefont {Lin}}, \bibinfo {author} {\bibfnamefont {Zhengtai}\ \bibnamefont {Liu}}, \bibinfo {author} {\bibfnamefont {Mao}\ \bibnamefont {Ye}}, \bibinfo {author} {\bibfnamefont {Yaobo}\ \bibnamefont {Huang}}, \bibinfo {author} {\bibfnamefont {Xiangyu}\ \bibnamefont {Meng}}, \bibinfo {author} {\bibfnamefont {Cheng}\ \bibnamefont {Zhang}}, \bibinfo {author} {\bibfnamefont {Taichi}\ \bibnamefont {Okuda}}, \bibinfo {author}
  {\bibfnamefont {Kenya}\ \bibnamefont {Shimada}}, \bibinfo {author} {\bibfnamefont {Shengtao}\ \bibnamefont {Cui}}, \bibinfo {author} {\bibfnamefont {Yue}\ \bibnamefont {Zhao}}, \bibinfo {author} {\bibfnamefont {Guang-Han}\ \bibnamefont {Cao}}, \bibinfo {author} {\bibfnamefont {Shan}\ \bibnamefont {Qiao}}, \bibinfo {author} {\bibfnamefont {Junwei}\ \bibnamefont {Liu}}, \ and\ \bibinfo {author} {\bibfnamefont {Chaoyu}\ \bibnamefont {Chen}},\ }\href {https://arxiv.org/abs/2407.19555} {\enquote {\bibinfo {title} {Crystal-symmetry-paired spin-valley locking in a layered room-temperature antiferromagnet},}\ } (\bibinfo {year} {2024}),\ \Eprint {http://arxiv.org/abs/2407.19555} {arXiv:2407.19555 [cond-mat.str-el]} \BibitemShut {NoStop}%
\bibitem [{\citenamefont {Kitamura}\ \emph {et~al.}(2024)\citenamefont {Kitamura}, \citenamefont {Daido},\ and\ \citenamefont {Yanase}}]{kitamura2024spin}%
  \BibitemOpen
  \bibfield  {author} {\bibinfo {author} {\bibfnamefont {Taisei}\ \bibnamefont {Kitamura}}, \bibinfo {author} {\bibfnamefont {Akito}\ \bibnamefont {Daido}}, \ and\ \bibinfo {author} {\bibfnamefont {Youichi}\ \bibnamefont {Yanase}},\ }\bibfield  {title} {\enquote {\bibinfo {title} {Spin-triplet superconductivity from quantum-geometry-induced ferromagnetic fluctuation},}\ }\href {https://journals.aps.org/prl/abstract/10.1103/PhysRevLett.132.036001} {\bibfield  {journal} {\bibinfo  {journal} {Physical Review Letters}\ }\textbf {\bibinfo {volume} {132}},\ \bibinfo {pages} {036001} (\bibinfo {year} {2024})}\BibitemShut {NoStop}%
\bibitem [{\citenamefont {Heinsdorf}(2025)}]{sitedecoupling2}%
  \BibitemOpen
  \bibfield  {author} {\bibinfo {author} {\bibfnamefont {Niclas}\ \bibnamefont {Heinsdorf}},\ }\bibfield  {title} {\enquote {\bibinfo {title} {Altermagnetic instabilities from quantum geometry},}\ }\href {\doibase 10.1103/PhysRevB.111.174407} {\bibfield  {journal} {\bibinfo  {journal} {Phys. Rev. B}\ }\textbf {\bibinfo {volume} {111}},\ \bibinfo {pages} {174407} (\bibinfo {year} {2025})}\BibitemShut {NoStop}%
\bibitem [{\citenamefont {Yao}\ \emph {et~al.}(2017)\citenamefont {Yao}, \citenamefont {Cai}, \citenamefont {Tong}, \citenamefont {Gong}, \citenamefont {Wang}, \citenamefont {Wan}, \citenamefont {Duan},\ and\ \citenamefont {Chu}}]{yao2017manipulation}%
  \BibitemOpen
  \bibfield  {author} {\bibinfo {author} {\bibfnamefont {Qun-Fang}\ \bibnamefont {Yao}}, \bibinfo {author} {\bibfnamefont {Jia}\ \bibnamefont {Cai}}, \bibinfo {author} {\bibfnamefont {Wen-Yi}\ \bibnamefont {Tong}}, \bibinfo {author} {\bibfnamefont {Shi-Jing}\ \bibnamefont {Gong}}, \bibinfo {author} {\bibfnamefont {Ji-Qing}\ \bibnamefont {Wang}}, \bibinfo {author} {\bibfnamefont {Xiangang}\ \bibnamefont {Wan}}, \bibinfo {author} {\bibfnamefont {Chun-Gang}\ \bibnamefont {Duan}}, \ and\ \bibinfo {author} {\bibfnamefont {JH}~\bibnamefont {Chu}},\ }\bibfield  {title} {\enquote {\bibinfo {title} {Manipulation of the large rashba spin splitting in polar two-dimensional transition-metal dichalcogenides},}\ }\href {https://journals.aps.org/prb/abstract/10.1103/PhysRevB.95.165401} {\bibfield  {journal} {\bibinfo  {journal} {Physical review B}\ }\textbf {\bibinfo {volume} {95}},\ \bibinfo {pages} {165401} (\bibinfo {year} {2017})}\BibitemShut {NoStop}%
\end{thebibliography}%


\newpage
~
\newpage

\newcommand{\<}{\langle}
\newcommand{\e}{\varepsilon}
\newcommand{\up}{\uparrow}
\newcommand{\down}{\downarrow}
\newcommand{\Up}{\Uparrow}
\newcommand{\Down}{\Downarrow}
\renewcommand{\>}{\rangle}
\renewcommand{\(}{\left(}
\renewcommand{\)}{\right)}
\renewcommand{\[}{\left[}
\renewcommand{\]}{\right]}
\renewcommand{\v}[1]{\boldsymbol{#1}} 
\newcommand{\dslash}{d \hspace{-0.8ex}\rule[1.2ex]{0.8ex}{.1ex}}
\renewcommand{\d}{\partial}
\newcommand{\del}{\nabla}
\renewcommand{\div}{\nabla\cdot}
\newcommand{\curl}{\nabla\times}
\newcommand{\eps}{\epsilon}
\newcommand{\p}{\parallel}
\newcommand{\U}{\mathcal{U}}

\appendix

\setcounter{figure}{0}
\setcounter{table}{0}
\setcounter{page}{1}
\makeatletter
\renewcommand{\thefigure}{S\arabic{figure}}
\renewcommand{\bibnumfmt}[1]{[S#1]}
\renewcommand{\citenumfont}[1]{S#1}

\onecolumngrid
\vspace{0.5cm}
\vspace{0.5cm}
\begin{center}
\bf \large End Matter
\end{center}
\vspace{0.5cm}
\twocolumngrid

{\em Spectrum of the effective Hamiltonian --} We analyze the spectrum of ${\tilde h}_1$ defined through Eqs.\ \eqref{h7} and \eqref{h10}. To facilitate analytical progress we neglect the band structure renormalization $\Omega_\bk$ and focus on the effect of proximity-induced SC gap encoded in $\Sigma_\bk$ which we write compactly as
\begin{equation}\label{h13}
  \Sigma_\bk=
  \begin{pmatrix}
   s_\bk & p_\bk \\
   p_\bk^\ast  &   s_\bk
    \end{pmatrix}.
\end{equation}
Here $s_\bk$ and $p_\bk$ denote the singlet and triplet component of the pairing matrix, respectively, defined by Eq.\ \eqref{h11}. The non-negative eigenvalues of the $4\times 4$ matrix Hamiltonian ${\tilde h}_1$ are then given by 
\begin{equation}\label{h14}
E_{\bk\pm}^2= \xi_\bk'^2+\eta_\bk^2+s_\bk^2+|p_\bk|^2\pm 2\sqrt{\xi_\bk'^2\eta_\bk^2+ \eta_\bk^2 s_\bk^2 + s_\bk^2 |p_\bk|^2 }.
\end{equation}

It is instructive to first consider the non-magnetic limit $\eta_\bk=0$ in which the spectrum simplifies and becomes
\begin{equation}\label{h15}
E_{\bk\pm}^2= \xi_\bk'^2 +(s_\bk \pm |p_\bk|)^2.
\end{equation}
We observe that singlet and triplet gaps are incompatible: they compete with each other and the spectrum in this limit becomes gapless whenever $s_\bk =\pm |p_\bk|$. We shall see that this property carries over to the altermagnetic case.

Other two instructive limits are when either $p_\bk$ or $s_\bk$ vanish. The corresponding spectra then become 
\begin{subequations}
\begin{align}
\label{h16a}
E_{\bk\pm} &= \sqrt{\xi_\bk'^2 +s_\bk^2} \pm \xi_\bk, \ \ \ \ \ \ \ \ \ \  
  {p_\bk=0 } 
  \\
  E_{\bk\pm} &= \sqrt{(\xi_\bk'\pm \eta_\bk)^2 +p_\bk^2}, \ \ \ \ \ \ \
  {s_\bk=0 } 
 \label{h16b}
\end{align}
\end{subequations}
The triplet component opens a gap regardless of the spin splitting magnitude whereas the singlet component only opens a gap when $|s_\bk|>|\eta_\bk|$. We note that based on Eq.\ \eqref{h11} it is always true that $|s_\bk|> |p_\bk|$. We may thus conclude that the singlet gap will be dominant close to the BZ diagonals where $|\eta_\bk|$ is small  while the triplet gap will dominate in regions of the BZ where $|\eta_\bk|$ is large. In view of Eq.\ \eqref{h15} we also expect the two regions to be separated by a nodal point where the excitation gap vanishes, which leads to the overall gap structure depicted in Fig.\ \ref{fig1}(b). These conclusions can be verified by numerically evaluating the full energy eigenvalues given in Eq.\ \eqref{h14}.

It is also possible to find the precise location of the nodal points. To this end we note that the expression for $E_{\bk-}^2$ given in Eq.\ \eqref{h14} becomes larger if we neglect the $\xi_\bk'^2\eta_\bk^2$ term under the square root sign. This leads to an inequality  
\begin{equation}\label{h17}
E_{\bk-}^2\leq \ \xi_\bk'^2+\left(|s_\bk| - \sqrt{\eta_\bk^2+ |p_\bk|^2}\right)^2 .
\end{equation}
Nodal points occur at crystal momenta $\bk$ where the right hand side vanishes, that is, when 
$\xi_\bk'=0$ and  $|s_\bk| = \sqrt{\eta_\bk^2+ |p_\bk|^2}$. The first condition, interestingly, places the nodal point on the non-magnetic Fermi surface of the ALM metal. On the Fermi surface we can approximate $\eta_\bk\simeq \eta_0^2 k_F^2\cos{2\alpha}$ where $\alpha$ is the angle defined in Fig.\ \ref{fig2}(a). Nodal points will thus be located at angles $\alpha$ satisfying 
\begin{equation}\label{h18}
\eta_0^2 k_F^2\cos{2\alpha} = \sqrt{s_\bk^2- |p_\bk|^2} \propto g^2.
\end{equation}
We deduce that for small $g$ nodal points will be close to the BZ diagonal but will move apart with increasing $g$ in accord with our numerical results for the complete bilayer system shown in Fig. \ref{fig2}(c). 

\end{document}